\documentclass[10pt,headings=small,a4paper,oneside,fleqn]{scrartcl}
\usepackage[utf8]{inputenc}
\usepackage[english]{babel}
\usepackage{placeins}
\usepackage[T1]{fontenc}
\usepackage{amsmath}
\usepackage{amsfonts}
\usepackage{amssymb}
\usepackage{graphicx}
\usepackage[left=2cm,right=2cm,top=2cm,bottom=2cm]{geometry}
\usepackage[version=3]{mhchem}
\usepackage{siunitx}
\usepackage{cite}
\usepackage{subcaption}
\usepackage[font=footnotesize,font=it,margin=10pt,format=plain]{caption}
\usepackage{multicol}
\usepackage{braket}
\usepackage{tikz}
\usepackage{float}
\usepackage[font=footnotesize,labelfont=bf]{caption}
\usepackage{multirow}
\usepackage{authblk}
\DeclareMathOperator{\sinc}{sinc}

\author[1,2]{\large Fabian Laudenbach$^{*}$}
\author[1]{\large Sebastian Kalista}
\author[1]{\large Michael Hentschel}
\author[2]{\large Philip Walther}
\author[1]{\large Hannes H\"{u}bel}
\affil[1]{\normalsize Optical Quantum Technology, Digital Safety \& Security Department,\protect \\AIT Austrian Institute of Technology GmbH, Donau-City-Str. 1, 1220 Vienna, Austria}
\affil[2]{\normalsize Quantum Optics, Quantum Nanophysics \& Quantum Information, Faculty of Physics, University of Vienna, Boltzmanngasse 5, 1090 Vienna, Austria \protect \\ \quad}
\affil[*]{\normalsize mail: fabian.laudenbach@ait.ac.at}

\title{A novel single-crystal~\&~single-pass source for polarisation- and colour-entangled photon pairs}
\date{}

\begin{document}

  \maketitle
\begin{abstract}
We demonstrate a new generation mechanism for polarisation- and colour-entangled photon pairs. In our approach we tailor the phase-matching of a periodically poled KTP crystal such that two downconversion processes take place simultaneously. Relying on this effect, our source emits entangled bipartite photon states, emerging intrinsically from a single, unidirectionally pumped crystal with uniform poling period. Its property of being maximally compact and luminous at the same time makes our source unique compared to existing photon-entanglement sources and is therefore of high practical significance in quantum information experiments.
\vspace{30pt}
\end{abstract}

\begin{multicols}{2}

Quantum entanglement is regarded as a key feature for new information technologies such as quantum key distribution \cite{ekert1991quantum}, dense coding \cite{bennet1992communication}, teleportation \cite{bouwmeester1997experimental} or one-way quantum computing \cite{knill2001ascheme}. In particular, most experiments and applications rely on photonic entanglement, taking advantage of long coherence times, good controllability and favourable detection methods. The most common generation technique relies on spontaneous parametric downconversion (SPDC) in  a $\chi^{(2)}$  nonlinear crystal. 

Up to this point photon-entanglement sources suffered from the compromise of being \emph{either} compact \emph{or} efficient. A very popular approach makes use of a $\beta$-barium borate (BBO) crystal to generate cones of down-converted photons whose lines of intersection are the spatial modes of polarisation-entangled photon states \cite{kwiat1995new, kurtsiefer2001high}. Although this source consists of a very simplistic setup\textemdash it uses only a single unidirectionally pumped crystal and is therefore very convenient to assemble\textemdash it allows for only moderate count rates since the generated photons are emitted spatially into a cone and only a small intersection is coupled into single-mode fibres. The rates can be improved by collapsing the cones into quasi-Gaussian modes propagating collinearly with the pump. However, in order to generate entanglement in this setup, \emph{two} crystals, arranged in a crossed fashion~\cite{kwiat1999ultrabright}, are needed. In addition, only a short propagation length is allowed as not to suffer from walk-off effects, thereby limiting the maximal production rate again. (As a side note, however, the non-collinear geometry of BBO-based sources has been demonstrated favourable for the generation of hyperentangled states~\cite{kwiat1997hyper, vallone2009hyperentanglement}. These states, among other interesting properties, allow for a convenient way to perform Bell-state analysis~\cite{kwiat1998embedded, liu2015complete}.)

The first truly collinear SPDC sources~\cite{tanzilli2001highly} appeared with the advent of quasi-phase-matching (QPM) in SPDC, allowing for a wider choice of wavelengths for the generated photon pairs, longer crystals, better mode matching with single-mode fibres and hence higher production rates. In order to generate (polarisation) entanglement, most existing QPM sources are either assembled in a single-pass~\&~double-crystal or single-crystal~\&~double-pass configuration. The most common double-crystal scheme is the Mach-Zehnder configuration \cite{yoshizawa2003generation,clausen2014source} where both crystals are pumped individually and the output is combined subsequently. Apart from higher costs and the elaborate effort of aligning two crystals, this type of setup also requires a stable phase relation to be kept between the two paths~\cite{herbauts2013demonstration}. Alternatively, the two crystals can be placed behind one another with the second crystal being rotated by $90^\circ$ with respect to the first one~\cite{pelton2004bright,huebel2007high}. In this realisation, the optimal focussing- and collection geometries cannot be satisfied for both crystals simultaneously, especially for long crystals and highly non-degenerate wavelengths of the produced photons. Due to the above disadvantages of the double-crystal configurations, a single-crystal~\&~double-pass geometry, based on a Sagnac interferometer, is widely used nowadays~\cite{shi2004generation, koenig2005efficient, kim2006phase, fedrizzi2007wavenength}. The remaining issues with this setup are a larger footprint and difficult alignment of the counter-propagating beams, especially for non-degenerate wavelengths \cite{hentschel2009three}. Recently polarisation entanglement has also been demonstrated in a single-crystal~\&~single-pass fashion where the periodic pattern of the QPM period is altered to support two different SPDC modes at the same time. This method is mainly applied in conjunction with optical waveguides where either the first and second section of the waveguide shows a different poling period \cite{suhara2009quasi}, or a more complex pattern in which the two different periods are interleaved \cite{herrmann2013post}. Apart from the time- and cost-extensive crystal manufacturing process, these promising approaches are at current state very sensitive to manufacturing imperfections and suffer from moderate count rates due to unsatisfying waveguide coupling.

Our new concept of creating entanglement combines most of the advantages of the previous schemes. At heart it uses a single nonlinear crystal with a uniform grating period for QPM which is also suitable for short-pulsed excitation. The setup offers high coupling efficiencies due to its collinear design and is still as compact and simplistic as a BBO source, requiring only one crystal pumped from one direction. The alignment process is therefore straightforward even for highly non-degenerate SPDC. By varying the pump wavelength, the emitted photon pairs can be wavelength-tuned to near degeneracy.

\section*{Collinear double-downconversion}

\begin{figure*}
\centering
\includegraphics[width=0.8\linewidth]{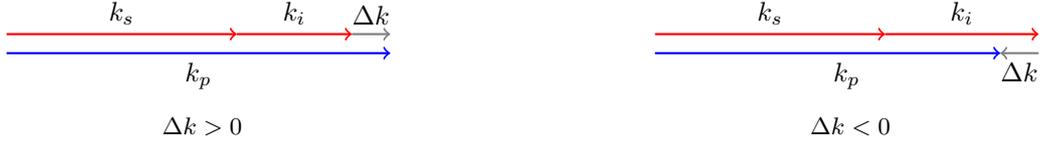}
\caption{The momentum-difference vector can be both positive and negative, thus requiring a positive or negative QPM order $m$ respectively.}
\label{mismatch}
\end{figure*}

Spontaneous parametric downconversion  is a quantum-mechanical process where a shortwave photon (usually referred to as \emph{pump}) decays into two longwave daughter photons (\emph{signal} and \emph{idler}) by interaction with a nonlinear medium. In general the three fields interfere destructively within the medium and no measurable output is generated. This is caused by a non-zero momentum-difference vector

\begin{align}
\Delta k = k_{p} - k_{s} - k_{i} = \frac{2\pi n_{p}}{\lambda_{p}} - \frac{2\pi n_{s}}{\lambda_{s}} - \frac{2\pi n_{i}}{\lambda_{i}} ,
\end{align}
where $n$ represent the respective refractive indices. Quasi-phase-matching  is a technique which allows for a positive energy transfer from pump to signal and idler fields throughout the medium. In this approach destructive interference of the three interacting fields is avoided by an alternating nonlinearity coefficient $d_{\text{eff}}$ of the material. In particular, ferro-electric poling of the crystal is used to toggle $d_{\text{eff}}$ between positive and negative values in integer multiples of a specific length along the crystal, the poling periodicity $\Lambda$. Consider a collinear process where pump, signal and idler each propagate along the crystal's $x$-axis. The amplitude for a downconversion in a crystal with length $L$ reads

\begin{align} \label{SPDCampl}
\ket{\Psi} = & \mathcal{N} L \int_{0}^{\infty} \int_{0}^{\infty} d_{\text{eff}}( \lambda_{s},\lambda_{i} ) \mu( \lambda_{p} ) \psi( \lambda_{s},\lambda_{i} ) \notag \\
& \times \hat{a}_{s}^{\dagger}( \lambda_{s} ) \hat{a}_{i}^{\dagger}( \lambda_{i} ) \; d \lambda_{s} \; d \lambda_{i} \ket{0} ,
\end{align}
where $\mathcal{N}$ is a normalisation constant, $d_{\text{eff}}$ is the effective nonlinearity coefficient for the specific process and $\hat{a}_{s,i}^{\dagger}$ are the respective creation operators. Moreover, $\mu$ is the \emph{pump envelope amplitude}, often represented by a Gaussian distribution:

\begin{equation}
\mu(\lambda_{p}) = \exp \left( \frac{-(\lambda_{p}-\lambda_{p0})^{2}}{2 \sigma^{2} } \right),
\end{equation}
where $\lambda_{p}=(1/\lambda_{s}+1/\lambda_{i})^{-1}$ by energy conservation and $\lambda_{p0}$ is the central pump wavelength. Finally, $\psi$ is the \emph{quasi-phase-matching amplitude}

\begin{equation}
\psi ( \lambda_{s},\lambda_{i} ) = e^{ i \Delta k_{m} L /2} \sinc \left( \frac{\Delta k_{m} L}{2} \right),
\end{equation}
with $\Delta k_{m}$ being the phase-mismatch vector represented as

\begin{equation} \label{Deltakm}
\Delta k_{m} = \Delta k - m \frac{2\pi}{\Lambda} ,
\end{equation}
where $m$ is an odd integer (positive or negative), referred to as the QPM order. Quasi-phase-matching is achieved when the amplitude $\psi$ is maximised. This is the case when the sinc function equals one or, equivalently, when $\Delta k_{m}=0$. From this requirement and from \eqref{Deltakm} follows the poling periodicity:

\begin{equation}
\Lambda = m \frac{2 \pi}{\Delta k} .
\end{equation}
Note that the momentum difference $\Delta k$ can be both positive and negative (Fig.~\ref{mismatch}). Accordingly the QPM order $m$ will be positive or negative respectively in order to preserve positivity of the crystal periodicity $\Lambda$. So more precisely, the quasi-phase-matching amplitude actually reads

\begin{align}
\psi ( \lambda_{s},\lambda_{i} ) & = \hspace{10pt} e^{ i \Delta k_{m_{+}} L /2} \sinc \left( \frac{\Delta k_{m_{+}} L}{2} \right) \notag \\
& \hspace{10pt} + e^{ i \Delta k_{m_{-}} L /2} \sinc \left( \frac{\Delta k_{m_{-}} L}{2} \right) \notag \\
& =: \psi_{+} ( \lambda_{s},\lambda_{i} ) + \psi_{-} ( \lambda_{s},\lambda_{i} ),
\end{align}
where the subscripts $+$ and $-$ correspond to a positive/negative QPM order $m$ in equation \eqref{Deltakm}. This modifies the SPDC amplitude \eqref{SPDCampl} to the state

\begin{align} \label{SPDCampl2}
\ket{\Psi} & = \hspace{10pt} \mathcal{N} L \int_{0}^{\infty} \int_{0}^{\infty} d_{\text{eff}} \mu \psi_{+} \hat{a}_{s}^{\dagger} \hat{a}_{i}^{\dagger} \; d \lambda_{s} \; d \lambda_{i} \ket{0} \notag \\
& \hspace{10pt} + \mathcal{N} L \int_{0}^{\infty} \int_{0}^{\infty} d_{\text{eff}} \mu \psi_{-} \hat{a}_{s}^{\dagger} \hat{a}_{i}^{\dagger} \; d \lambda_{s} \; d \lambda_{i} \ket{0} \\ \notag
& =: \alpha \ket{\Psi_{+}} + \beta \ket{\Psi_{-}} ,
\end{align}
where $|\alpha|^{2} + |\beta|^{2} = 1$. While in most experimental setups only one of the two terms contributes to the amplitude there are specific configurations (crystal type, wavelengths, polarisations) where neither of both terms can be neglected. It is important to note that in this case a single periodically poled crystal with a given grating $\Lambda$ and pumped by a laser with given centre wavelength $\lambda_{p0}$ can be used to phase-match up to \emph{two} different SPDC processes with distinct wavelength pairs being generated\textemdash provided the respective momentum-difference vectors $\Delta k$ carry the same length but different signs. This property, collinear double-downconversion (CDDC), is the key to our novel entangled-photon source.

\section*{Intrinsic entanglement}

\begin{figure*}
\centering
\includegraphics[width=\linewidth]{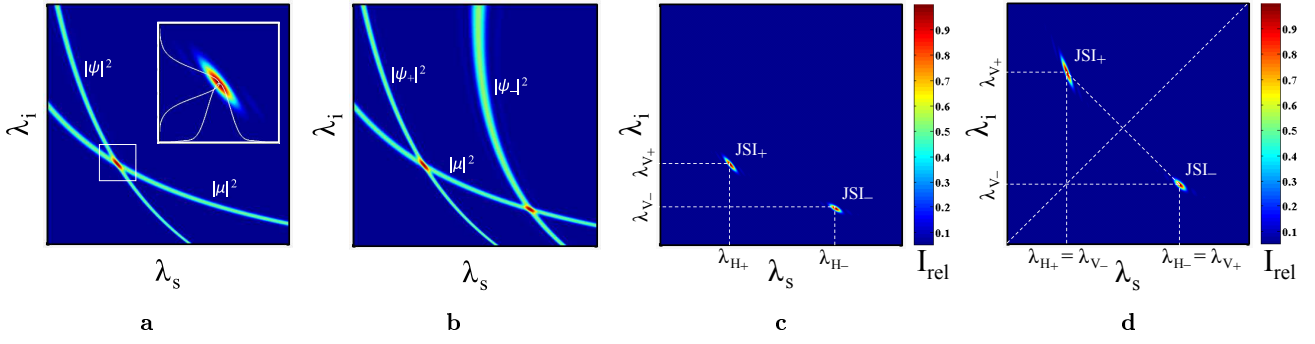}
\caption{Depiction of the joint spectral intensity. Figure~\textbf{a} illustrates how the intersection of the pump intensity $|\mu|^{2}$ and the phase-matching intensity $|\psi|^{2}$ shapes the joint spectral intensity distribution of signal and idler. CDDC occurs in particular configurations where $|\mu|^{2}$ intersects with both components of the phase-matching intensity, namely $|\psi_{+}|^{2}$ and $|\psi_{-}|^{2}$, as illustrated in Figure~\textbf{b}. This means that one crystal with a given periodicity $\Lambda$, pumped by a laser with given centre wavelength $\lambda_{p0}$, can evoke \emph{two} sub-JSI's, generating photons with a total of \emph{four} different wavelengths, as shown in Figure~\textbf{c}. In addition, as depicted in Figure~\textbf{d}, under careful choice of the setup, the two sub-JSI's can be composed such that the corresponding wavelengths become interchangeable, resulting in an entangled bipartite state.}
\label{JSIgraphs}
\end{figure*}

The \emph{joint spectral intensity} (JSI) of an SPDC process is the probability distribution function of specific signal and idler wavelength pairs being emitted. It is represented as the square over the joint spectral amplitude (JSA) which is the product of pump- and phase-matching envelope amplitude:

\begin{equation}
\text{JSI}=|\text{JSA}|^{2}=|\mu( \lambda_{p} ) \psi( \lambda_{s},\lambda_{i} )|^{2} ,
\end{equation}
with $\lambda_{p}=(1/\lambda_{s}+1/\lambda_{i})^{-1}$. Graphically, when intensity over signal and idler wavelength is plotted, the JSI can be understood as the intersection of the envelope intensities $|\mu|^{2}$ and $|\psi|^{2}$, as illustrated in Figure~\ref{JSIgraphs}a. When a periodically poled crystal allows for phase-matching of both positive and negative $\Delta k$ in a given wavelength range, the pump intensity $|\mu|^{2}$ intersects with \emph{two} constituents of $\psi(\lambda_{s},\lambda_{i})$, namely $|\psi_{+}|^{2}$ and $|\psi_{-}|^{2}$, as shown in Figure~\ref{JSIgraphs}b. In this case the setup will generate photons with four distinct wavelengths: one pair with $\lambda_{s_{+}}$, $\lambda_{i_{+}}$ and joint spectral intensity $\text{JSI}_{+}$ and another one with $\lambda_{s_{-}}$, $\lambda_{i_{-}}$ and $\text{JSI}_{-}$ (Fig.~\ref{JSIgraphs}c).

Consider now a type-II downconversion where signal and idler radiation are orthogonally polarised and define, without loss of generality, the signal photons to be polarised horizontally. Under careful configuration of the setup, the two centre wavelengths of one pair can be brought to coincide with those of the respective counterparts of the other, orthogonally polarised, pair, hence

\begin{subequations}
\begin{align}
\lambda_{H_{+}} & = \lambda_{V_{-}}, \\
\lambda_{V_{+}} & = \lambda_{H_{-}},
\end{align}
\end{subequations}
as illustrated in Figure~\ref{JSIgraphs}d. When the wavelengths of the two pairs cannot be told apart, entanglement emerges from the lost information whether a generated pair originates from $\ket{\Psi_{+}} $ or $\ket{\Psi_{-}} $. So when the two joint spectral distributions become interchangeable in wavelength, the SPDC amplitude \eqref{SPDCampl2} becomes entangled in polarisation and wavelength. When we label the shortwave daughter photons as blue ($B$) and the longwave ones as red ($R$), the state can (in a simplified form) be written as

\begin{align}
\ket{\Psi} & = \alpha \ \hat{a}^{\dagger}_{HB} \hat{a}^{\dagger}_{VR} \ket{0} + \beta \ \hat{a}^{\dagger}_{HR} \hat{a}^{\dagger}_{VB} \ket{0} \notag \\
& =: \alpha \ket{H V} \otimes \ket{B R} + \beta \ket{H V}\otimes  \ket{R B} ,
\end{align}
with all creation operators acting on the same spatial mode due to the collinearity of the downconversion. Depending on experimental requisition this state can be reduced to both a \emph{polarisation}-entangled state

\begin{equation} \label{polent}
\ket{\Psi}=\left( \alpha \ket{H_{1} V_{2}} + \beta \ket{V_{1} H_{2}} \right) \otimes \ket{B_{1} R_{2}}
\end{equation}
or \emph{frequency}-entangled state

\begin{equation} \label{freent}
\ket{\Psi}=\left( \alpha \ket{B_{1} R_{2}} + \beta \ket{R_{1} B_{2}} \right) \otimes \ket{H_{1} V_{2}},
\end{equation}
where the photons are spatially separated into mode 1 and 2 by a dichroic mirror \eqref{polent} or a polarising beamsplitter \eqref{freent}. The amplitudes $\alpha$ and $\beta=|\beta| e^{i\theta}$ are complex numbers which fulfil the normalisation condition $|\alpha|^{2}+|\beta|^{2}=1$. Note that in general the effective nonlinearity $d_{\text{eff}}$ of the two superimposed SPDC processes is not of equal magnitude, so neither will be $|\alpha|$ and $|\beta|$. In addition, the two JSI distributions might come in different spectral bandwidths which has to be compensated using sufficiently narrow bandpass filters in order to avoid spectral distinguishability. This will further influence the magnitude of the two amplitudes. 

\section*{Numerical investigation of CDDC-entanglement in ppKTP}

\begin{figure*}
    \centering
\includegraphics[width=0.8\linewidth]{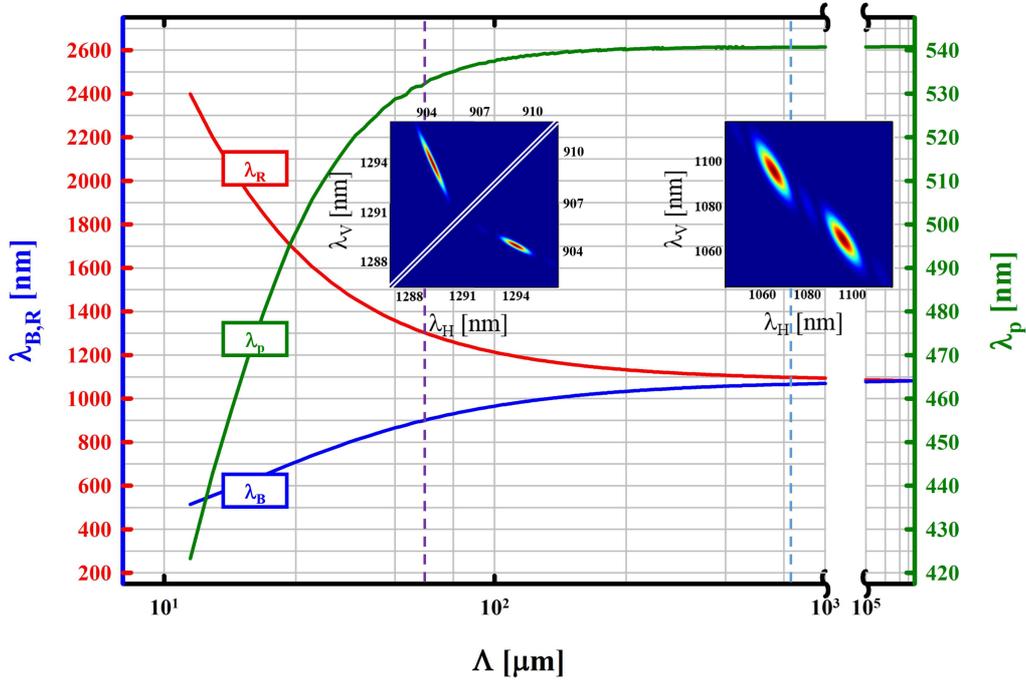}
\caption{Possible configurations allowing for intrinsic entanglement of photons generated by CDDC in \text{ppKTP}. The blue and red lines represent the short- and longwave photons respectively (each coming in both polarisations $H$ and $V$); the green line represents the wavelength of the pump laser. The dashed purple line and corresponding inset graph highlight our specific experimental setup with $\lambda_{p0}=\SI{532.3}{\nano\metre}$, $\lambda_{B}=\SI{904.3}{\nano\metre}$ and $\lambda_{R}=\SI{1293.9}{\nano\metre}$. With increasing crystal periodicity $\Lambda$ the wavelengths $\lambda_{B}$ and $\lambda_{R}$ are approaching each other  which gives rise to the opportunity of creating entangled photon pairs with similar bandwidths and group velocities, as depicted by the right inset graph over the light blue dashed line. The calculations were performed assuming a crystal temperature of $T=\SI{50}{\degreeCelsius}$. However, the actual temperature required for accurate matching of horizontal and vertical wavelength may vary due to imperfection of the used dispersion equations \cite{fan1987second,fradkin1999tunable,emanueli2003temperature}.}
\label{allconfigs}
\end{figure*}

Using our software `QPMoptics' \cite{laudenbach2016qpmoptics}, we discovered a large variety of configurations allowing for generation of intrinsic entanglement using CDDC in periodically poled potassium titanyl phosphate (\text{KTiOPO}$_{4}$, \text{ppKTP}). These configurations are depicted in Figure~\ref{allconfigs}. The figure shows that the generated wavelengths $\lambda_{B}$ and $\lambda_{R}$ approach each other as the crystal periodicity $\Lambda$ gets longer. This brings the advantage of equal group velocities in the two SPDC processes, since differences in the wavelength dispersion can be neglected. Moreover, due to overlapping bandwidths for photons with same polarisation (see right inset in Figure~\ref{allconfigs}), no measures such as bandpass-filtering have to be taken in order to preserve coherent overlap of the two JSI's. This gives rise to the possibility of a highly efficient colour-entanglement source with signal and idler in the vicinity of $\lambda_{s,i}\sim\SI{1}{\micro\metre}$. Since no filtering is required, this source would be very promising in terms of brightness and visibility. However, the  degeneracy wavelength lies at around 1100~nm, invisible to common avalanche photo diodes, and would require superconducting nanowire detectors to be counted efficiently. For the first demonstration of intrinsic entanglement from CDDC we therefore decided on highly non-degenerate polarisation-entangled photon pairs at roughly $\lambda_{B}\sim\SI{900}{\nano\metre}$ and $\lambda_{R}\sim\SI{1300}{\nano\metre}$ which can be detected efficiently with available Silicon and InGaAs photo diodes respectively.

\section*{Experiment}

\begin{figure*}
\centering
\includegraphics[width=0.8\linewidth]{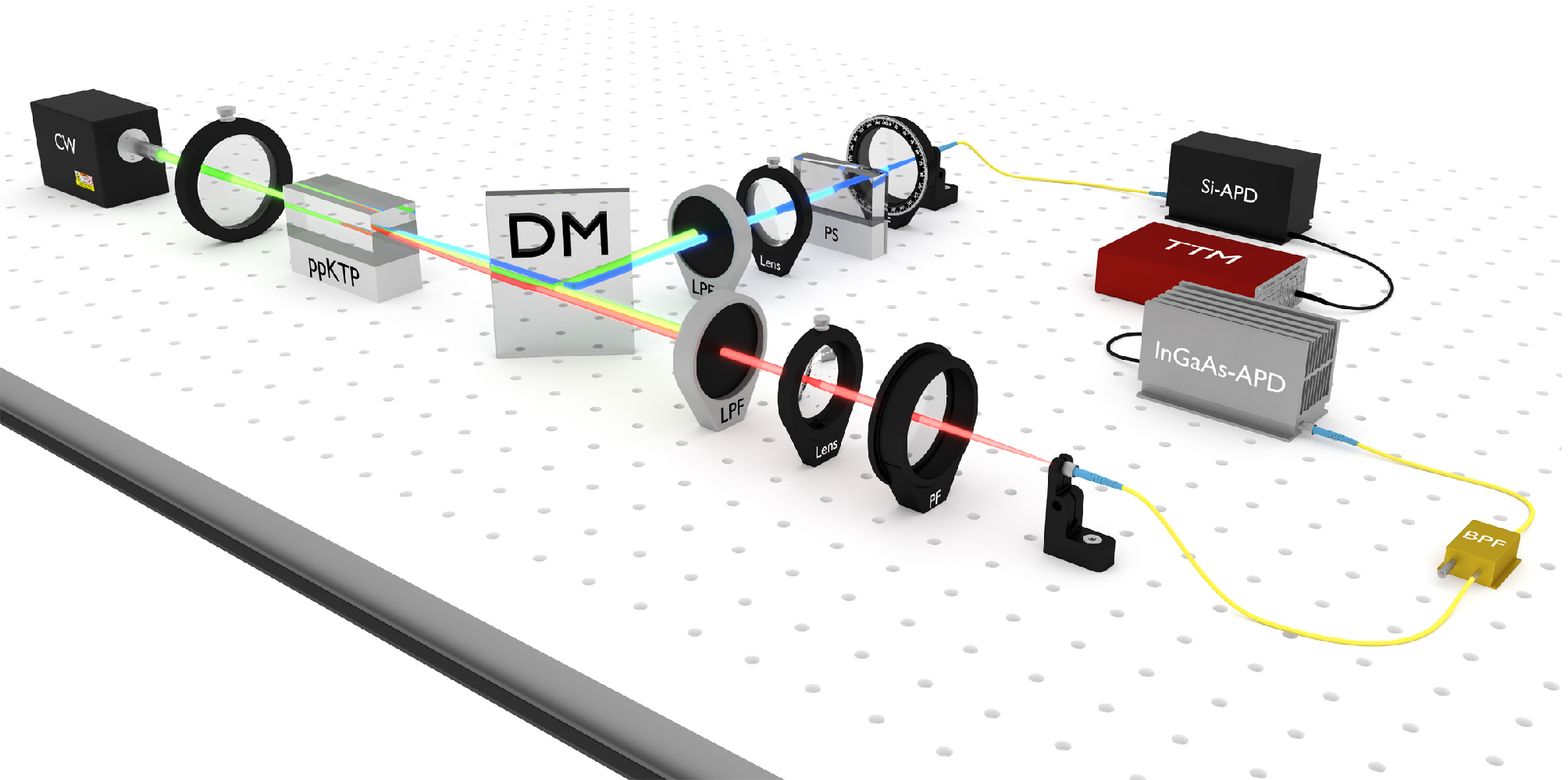}
\caption{Experimental setup. A $\SI{532.3}{\nano\metre}$ continuous-wave laser pumps a ppKTP crystal with periodicity $\Lambda=\SI{63.1}{\micro\metre}$, length $L=\SI{10}{\milli\metre}$ and temperature $T=\SI{60.0}{\degreeCelsius}$ to generate photon pairs at $\lambda_{B}=\SI{904.3}{\nano\metre}$ and $\lambda_{R}=\SI{1293.9}{\nano\metre}$, both in horizontal and vertical polarisation. A dichroic mirror (DM) separates the generated pairs according to their wavelength. Longpass filters (LPF) in both arms are used to remove the pump light. A phase shifter (PS), realised by a pair of calcite wedges in the $\SI{904.3}{\nano\metre}$ arm, allows us to tune the phase angle $\theta$ and to compensate for the different relative group delays of the two superimposed photon pairs. After going through a rotatable polarisation filter (PF) the photons are coupled into single-mode fibres. A fibre-coupled, tunable bandpass filter (BPF) of $\SI{0.9}{\nano\metre}$ width, limits the bandwidths of the horizontal and vertical $\SI{1293.9}{\nano\metre}$ photons. We used a silicon avalanche photo diode (Si-APD, $\eta_{\text{det}}\approx 0.38$) and an indium gallium arsenide (InGaAs) APD ($\eta_{\text{det}}\approx 0.12$) to detect the $\SI{904.3}{\nano\metre}$ and the $\SI{1293.9}{\nano\metre}$ photons respectively. A time-tagging module (TTM) was used to count the coincidences within a time window of $\SI{2}{\nano\second}$. (Image \textcopyright\ Alexander Sanchez de la Cerda)}
\label{setup}
\end{figure*}

\begin{figure*}
\centering
\includegraphics[width=\linewidth]{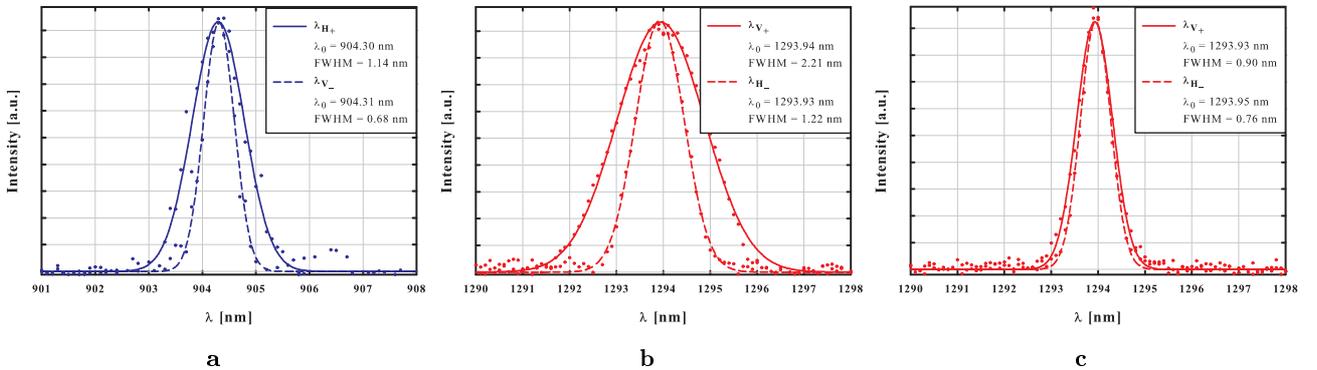}
\caption{Wavelength distribution of the down-converted photons. The solid (dashed) line represents photons from JSI$_{+}$ (JSI$_{-}$).  Figure~\textbf{a} depicts the spectrum of the horizontally and vertically polarised shortwave photon. The spectra of the longwave photons are shown in \textbf{b}. Spectral filtering of the longwave photons matches the two spectra, as depicted in \textbf{c}.}
\label{spectra}
\end{figure*}

The experimental setup is depicted in Figure~\ref{setup}. A 10 mm long, periodically poled KTP crystal was pumped with a $\lambda_{p0}=\SI{532.3}{\nano\metre}$ continuous-wave laser, generating photon pairs at $\lambda_{B}=\SI{904.3}{\nano\metre}$ and $\lambda_{R}=\SI{1293.9}{\nano\metre}$. The photons were separated according to their wavelength using a dichroic mirror, coupled into single-mode fibres and detected by avalanche photo diodes. Even though the centre-wavelengths of the generated photon pairs can be matched very accurately, entanglement visibility is undermined by spectral distinguishability due to different bandwidths of the two JSI's. This distinguishability could be overcome by insertion of a narrow tunable bandpass filter (BPF) in the arm of the longwave photon ($\lambda_{R}$). The spectra of the down-converted photons before and after bandpass filtering are illustrated in Figure~\ref{spectra}. For our experiment it turned out that the process with higher nonlinearity (and therefore SPDC amplitude) is also the one with greater spectral bandwidth. Therefore, insertion of a bandpass filter reduced the greater one of the two amplitudes and matched the coefficients $|\alpha|$ and $|\beta|$ to each other such that we ended up with an almost maximally entangled state:

\begin{equation} \label{Psi}
\ket{\Psi} \approx \frac{1}{\sqrt{2}} \left( \ket{H_{1} V_{2}} + e^{i\theta} \ket{V_{1} H_{2}} \right).
\end{equation}
Due to different group velocities within the crystal, the relative time delays between the shortwave and longwave photon differ, according to whether the pair originates from $\ket{\Psi_{+}}$ or $\ket{\Psi_{-}}$ (see~Table~\ref{datatable}), therefore diminishing coherent superposition of the two SPDC processes. This effect could be compensated using birefringent calcite wedges. The very same wedges were also used for tuning of the phase angle $\theta$ of the entangled state \eqref{Psi}.

\begin{figure*}
\centering
\includegraphics[width=0.7\linewidth]{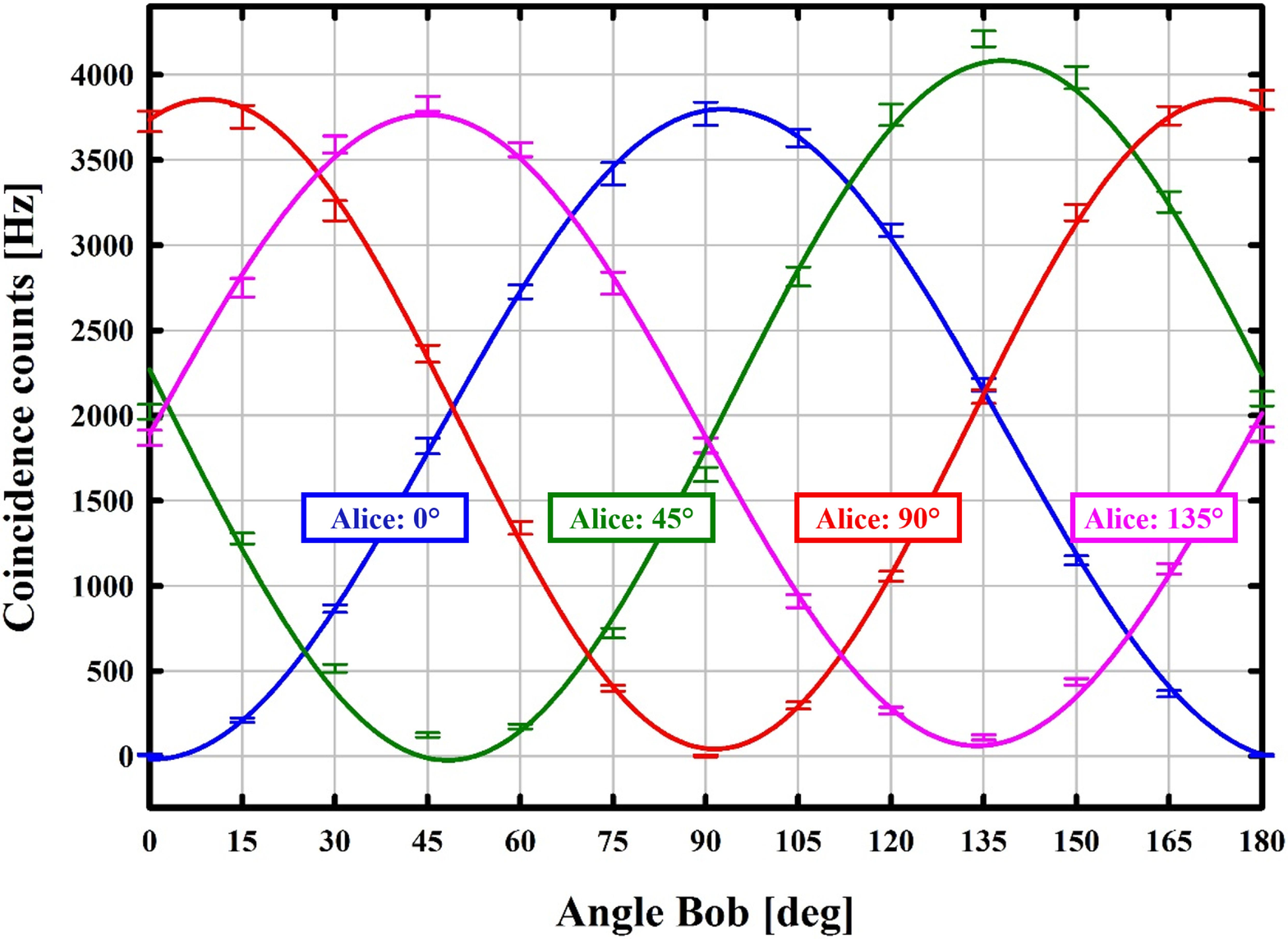}
\caption{Measured coincidence counts and sinusoidal fit (solid lines) with respect to polariser settings. The visibilities in the H/V-basis (blue and red) and the diagonal basis (green and pink) are $V_{H/V}=0.999(4)$ and $V_{+/-}=0.971(6)$ respectively. The measurements were taken at a pump power of $P=\SI{8.3}{mW}$ and are averaged over approximately one minute; the error bars represent one standard deviation $\sigma$ of the mean value.}
\label{correlations}
\end{figure*}

Using single-photon counters in both arms, we measured a coincidence rate of $\SI{1200}{counts/s/mW}$. Taking into account coupling- and detection losses, this corresponds to a pair-generation rate of approximately $5.5 \times 10^{5}~\text{counts/s/mW}$ and a spectral brightness of $3.4 \times 10^{6}~\text{counts/s/mW/THz}$. (To give a comparison with up-to-date single-crystal \& single-pass entanglement sources, the\textemdash to our knowledge\textemdash brightest BBO-based source~\cite{kurtsiefer2001high} yielded a measured coincidence rate of up to $\SI{900}{counts/s/mW}$, a pair-generation rate of $1.1 \times 10^{4}~\text{counts/s/mW}$ and a spectral brightness of $4.5 \times 10^{3}~\text{counts/s/mW/THz}$.) We measured polarisation correlations in the H/V basis ($0^{\circ}$ and $90^{\circ}$) and in the diagonal basis ($\pm 45^{\circ}$), depicted in Figure~\ref{correlations}. After subtraction of the background coincidences, we achieved visibilities of $V_{H/V}=0.999(4)$ and $V_{+/-}=0.971(6)$ respectively. The average visibility is therefore $V_{\text{avg}}=0.985(4)$ which exceeds the non-locality threshold of $1/\sqrt{2} \approx 0.71$ by more than 71 standard deviations. (Before subtraction of the background we measured $V_{H/V}=0.985(3)$, $V_{+/-}=0.950(5)$ and therefore $V_{\text{avg}}=0.967(3)$.) We measured a CHSH parameter \cite{clauser1969proposed} of $S=2.817(22)$ ($S=2.778(21)$ before subtraction of the background pairs), therefore violating the CHSH inequality by approximately 38 standard deviations. After removal of the tunable bandpass filter, the coincidence rate could be increased to $\SI{1970}{counts/s/mW}$ under the cost of a lower visibility in the diagonal basis: $V_{+/-}=0.733(12)$, therefore $V_{\text{avg}}=0.865(6)$. The heralding efficiency of the longwave photon amounts to $\SI{45.1}{\%}$ which is a drastic advance compared to other highly non-degenerate entangled-photon sources (e.g.\ $~20~\%$ \cite{huebel2007high} or $10~\%$ \cite{hentschel2009three}).

\section*{Conclusion and outlook}

We presented the theoretical concept and first experimental demonstration of a novel collinear entangled-photon source, consisting of a single unidirectionally pumped nonlinear crystal. Using the collinear double-downconversion effect in periodically poled crystals, this concept can be used to generate both, polarisation- and colour-entangled photon states. Numerical evaluations revealed a broad range of wavelength pairs this type of source could generate. In our particular experimental setup we successfully generated polarisation-entangled pairs of \SI{904.3}{\nano\metre} and \SI{1293.9}{\nano\metre} wavelength, achieving coincidence rates of $\SI{1200}{counts/s/mW}$ and an average visibility of $V_{\text{avg}}=0.985(4)$. However, we found that pairs in the vicinity of $\lambda\sim \SI{1}{\micro\metre}$ would feature even better spectral and temporal indistinguishability due to coinciding bandwidths and group velocities in the medium. This would enhance the entanglement visibility and allow for significantly higher count rates, since no bandpass-filtering would be required to match the bandwidths. Moreover, we numerically investigated four other crystals in the \emph{mm2} point group and discovered a large number of additional promising implementations of our method (Laudenbach \textit{et al.} Numerical investigation of photon-pair generation in periodically poled \textit{M}TiO\textit{X}O$_{4}$ (\textit{M} = K, Rb, Cs; \textit{X} = P, As) crystals. \textit{Manuscript under preparation}).

Due to its collinear and very compact design our novel source will set new standards in terms of scalability as it is required for example in multi-photon experiments.

\section*{Methods}

\begin{table*}
\centering
\begin{tabular}{|c|c|c|c|c|c|c|c|}
\hline 
$m$ & $\lambda$ [nm] &  Pol & $\Delta \lambda_{\text{sim}}$ [nm] & $\Delta \lambda_{\text{meas}}$ [nm] & $\tau_{\text{coh}}$ [ps] & GD [ps/mm] & $\Delta t_{\text{max}}$ [ps] \\ 
\hline \hline
\multirow{2}{*}{$+1$}  & 904.3 & H & 1.1 & 1.1 & 2.5 & 6.0 & \multirow{2}{*}{$-2.2$} \\ 
& 1293.9 & V & 2.2 & 2.2 $\stackrel{\text{{\tiny BPF}}}{\longrightarrow}$ 0.90 & 2.5 $\stackrel{\text{{\tiny BPF}}}{\longrightarrow}$ 6.2 & 6.2 &  \\ 
\hline
\multirow{2}{*}{$-1$} & 904.3 & V & 0.60 & 0.68 & 3.9 & 6.3 & \multirow{2}{*}{4.0} \\ 
& 1293.9 & H & 1.2 & 1.2 $\stackrel{\text{{\tiny BPF}}}{\longrightarrow}$ 0.76 & 4.6 $\stackrel{\text{{\tiny BPF}}}{\longrightarrow}$ 7.0 & 5.9 &  \\
\hline
\end{tabular}
\caption{Centre wavelength $\lambda$, polarisation, spectral bandwidth $\Delta \lambda$ (numerically simulated and measured), coherence time $\tau_{\text{coh}}$ and group delay GD for each photon of the two pairs. The bandpass-filtering (BPF) is required to match the longwave photons' bandwidths for the sake of spectral indistinguishability. $\Delta t_{\text{max}}$ represents the maximal time delay between photons of one generated pair, obtained by comparing their total group delays in a 10~mm crystal. One can see that in one SPDC process the shortwave photon can leave the crystal up to 2~ps earlier whereas in the other one the shortwave photon may be delayed by up to 4~ps with respect to the longwave one. The measured discrepancy in the filtered bandwidths (0.90 vs. 0.76 nm) stems from the resolution limit of our spectrometer (0.1 nm).}
\label{datatable}
\end{table*}

A detailed description of the experimental setup is provided in the caption of Figure~\ref{setup}. The wavelength spectra of our generated photons (Fig.~\ref{spectra}) were measured using an optical spectrum analyser (OSA) in conjunction with our single-photon detectors. In its normal operation mode, the OSA measures the wavelength-resolved optical power with a built-in detector which, however, is not sensitive enough to resolve single photons. We therefore operated the OSA as a monochromator by replacing the internal detector with a fibre output that we connected to an external single-photon detector. A self-written Labview program was used to control the monochromator's diffraction grating and log the count rate with respect to the coupled wavelength.

Before planning and conducting this experiment, we performed a thorough numerical search for possible realisations of our novel concept. Apart from type-II downconversion in periodically poled KTP, we initially investigated two other popular nonlinear materials, namely lithium niobate (\text{LiNbO}$_{3}$, \text{ppLN}) and lithium tantalate (\text{LiTaO}$_{3}$, \text{ppLT}) but were unable to find experimentally feasible setups using these media. However, in recent numerical investigations, we discovered a large variety of possible implementations of our novel concept using other nonlinear materials in the \emph{mm2} point group (isomorphic to KTP). These results help to make new favourable wavelength configurations accessible (Laudenbach \textit{et al.} Numerical investigation of photon-pair generation in periodically poled \textit{M}TiO\textit{X}O$_{4}$ (\textit{M} = K, Rb, Cs; \textit{X} = P, As) crystals. \textit{Manuscript under preparation}).

The experiment was designed using our SPDC-simulation software QPMoptics~\cite{laudenbach2016qpmoptics}. Table~\ref{datatable} provides a direct comparison of the numerically simulated spectra with the measured ones as well as the photons' individual group delays and relative time delays.

In order to obtain an estimate on the pair-generation rate PR, we took into account that the measured single-count rates SR of the bluer (B) and redder (R) arm as well as the measured coincidence rate CR are attenuated by imperfect coupling efficiency $\mu$ and detection efficiency $\eta$:

\begin{subequations}
\begin{align}
\text{SR}_{B} & =\text{PR} \cdot \mu_{B} \cdot \eta_{B}, \\
\text{SR}_{R} & =\text{PR} \cdot \mu_{R} \cdot \eta_{R}, \\
\text{CR} & =\text{PR} \cdot \mu_{B} \cdot \mu_{R} \cdot \eta_{B} \cdot \eta_{R} .
\end{align}
\end{subequations}
This allows us to express the coupling efficiencies as follows:

\begin{subequations}
\begin{align}
\mu_{B} & =\frac{\text{CR}}{\eta_{B} \cdot \text{SR}_{R}}, \\
\mu_{R} & =\frac{\text{CR}}{\eta_{R} \cdot \text{SR}_{B}}.
\end{align}
\end{subequations}
The pair-generation rate is then

\begin{align}
\text{PR} = \frac{\text{CR}}{\mu_{B} \cdot \mu_{R} \cdot \eta_{B} \cdot \eta_{R}} = \frac{\text{SR}_{B} \cdot \text{SR}_{R}}{\text{CR}} .
\end{align}
The spectral brightness is the pair-generation rate per optical pump power $P$ per photon bandwidth $B$:

\begin{subequations}
\begin{align}
\text{Brightness}_{R} & = \frac{\text{PR}}{ P \cdot B_{R}} = \frac{\text{SR}_{B} \cdot \text{SR}_{R}}{\text{CR} \cdot P \cdot B_{R}}, \\
\text{Brightness}_{B} & = \frac{\text{SR}_{B} \cdot \text{SR}_{R}}{\text{CR} \cdot P \cdot B_{B}}. 
\end{align}
\end{subequations}

\section*{References}

\begingroup
\renewcommand{\section}[2]{}

\endgroup

\section*{Acknowledgements}

We gratefully acknowledge invaluable help from Florian Stierle and Alexander Sanchez de la Cerda and Christoph Pacher. This work was funded by the Austrian Research Promotion Agency (Österreichische Forschungsförderungsgesellschaft, FFG) through KVQ (No. 4642983). Moreover, we acknowledge support from the European Commission, via EQUAM (No. 323714), PICQUE (No. 608062), GRASP (No. 613024), QUCHIP (No. 641039) and the Austrian Science Fund (FWF) through START (Y585-N20).

\section*{Author contributions}

F.L. conceived the concept and theoretical foundations of the experiment and performed the numerical search for possible realisations. H.H. operated as the project leader and, together with M.H., developed the schedule, methodology and detailed procedure of the experiment. Moreover, M.H. contributed to the experimental setup. S.K. assembled the source and performed most measurements. P.W. contributed the theoretical analysis of the polarisation- and frequency-entangled state. The manuscript draft was written by F.L. and all co-authors contributed to the final version.

\section*{Competing financial interests}

F.L. and the AIT will be holding a share of a possible commercial distribution of the SPDC-simulation software `QPMoptics' by a third-party retailer.

\end{multicols}


\begin{thebibliography}{99}

\bibitem{ekert1991quantum} Ekert, A. K. Quantum cryptography based on Bell’s theorem. \textit{Phys. Rev. Lett.} \textbf{67,} 661 (1991).

\bibitem{bennet1992communication} Bennett, C. H. \& Wiesner, S. J. Communication via one-and two-particle operators on Einstein-Podolsky-Rosen states. \textit{Phys. Rev. Lett.} \textbf{69,} 2881 (1992).

\bibitem{bouwmeester1997experimental} Bouwmeester, D. \emph{et al}. Experimental quantum teleportation. \textit{Nature} \textbf{390,} 575--579 (1997).

\bibitem{knill2001ascheme} Knill, E., Laflamme, R. \& Milburn, G. J. A scheme for efficient quantum computation with linear optics. \textit{Nature} \textbf{409,} 46--52 (2001).

\bibitem{kwiat1995new} Kwiat, P. G. \emph{et al}. New high-intensity source of polarization-entangled photon pairs. \textit{Phys. Rev. Lett.} \textbf{75,} 4337 (1995).

\bibitem{kurtsiefer2001high} Kurtsiefer, C., Oberparleiter, M., \& Weinfurter, H. High-efficiency entangled photon pair collection in type-II parametric fluorescence. \textit{Phys. Rev. A} \textbf{64,} 023802 (2001).

\bibitem{kwiat1999ultrabright} Kwiat, P. G., Waks, E., White, A. G., Appelbaum, I. \& Eberhard, P. H. Ultrabright source of polarization-entangled photons. \textit{Phys. Rev. A} \textbf{60,} R773 (1999).

\bibitem{kwiat1997hyper} Kwiat, P. G. Hyper-entangled states. \textit{J. Mod. Optic}, \textbf{44,} 2173-2184 (1997).

\bibitem{vallone2009hyperentanglement} Vallone, G., Ceccarelli, R., De Martini, F., \& Mataloni, P. Hyperentanglement of two photons in three degrees of freedom. \textit{Phys. Rev. A} \textbf{79,} 030301 (2009).

\bibitem{kwiat1998embedded} Kwiat, P. G., \& Weinfurter, H. Embedded Bell-state analysis. \textit{Phys. Rev. A} \textbf{58,} R2623 (1998).

\bibitem{liu2015complete} Liu, Q., Wang, G. Y., Ai, Q., Zhang, M., \& Deng, F. G. Complete nondestructive analysis of two-photon six-qubit hyperentangled Bell states assisted by cross-Kerr nonlinearity. \textit{Sci. Rep.} \textbf{6,} 22016 (2015).

\bibitem{tanzilli2001highly} Tanzilli, S. \emph{et al}. Highly efficient photon-pair source using periodically poled lithium niobate waveguide. \textit{Electron. Lett.} \textbf{37,} 26--28 (2001).

\bibitem{yoshizawa2003generation} Yoshizawa, A., Kaji, R. \& Tsuchida, H. Generation of polarization-entangled photon pairs at 1550 nm using two PPLN waveguides. \textit{Electron. Lett.} \textbf{39,} 621--622 (2003).

\bibitem{clausen2014source} Clausen, C. \emph{et al}. A source of polarization-entangled photon pairs interfacing quantum memories with telecom photons. \textit{New J. Phys.} \textbf{16,} 093058 (2014).

\bibitem{herbauts2013demonstration} Herbauts, I., Blauensteiner, B., Poppe, A., Jennewein, T. \& Huebel, H. Demonstration of active routing of entanglement in a multi-user network. \textit{Opt. Express} \textbf{21,} 29013--29024 (2013).

\bibitem{pelton2004bright} Pelton, M. \emph{et al}. Bright, single-spatial-mode source of frequency non-degenerate, polarization-entangled photon pairs using periodically poled KTP. \textit{Opt. Express} \textbf{12,} 3573--3580 (2004).

\bibitem{huebel2007high} Hübel, H. \emph{et al}. High-fidelity transmission of polarization encoded qubits from an entangled source over 100 km of fiber. \textit{Opt. Express} \textbf{15,} 7853--7862 (2007).

\bibitem{shi2004generation} Shi, B. S. \& Tomita, A. Generation of a pulsed polarization entangled photon pair using a Sagnac interferometer. \textit{Phys. Rev. A} \textbf{69,} 013803 (2004).

\bibitem{koenig2005efficient} König, F., Mason, E. J., Wong, F. N. \& Albota, M. A. Efficient and spectrally bright source of polarization-entangled photons. \textit{Phys. Rev. A} \textbf{71,} 033805 (2005).

\bibitem{kim2006phase} Kim, T., Fiorentino, M. \& Wong, F. N. Phase-stable source of polarization-entangled photons using a polarization Sagnac interferometer. \textit{Phys. Rev. A} \textbf{73,} 012316 (2006).

\bibitem{fedrizzi2007wavenength} Fedrizzi, A., Herbst, T., Poppe, A., Jennewein, T. \& Zeilinger, A. A wavelength-tunable fiber-coupled source of narrowband entangled photons. \textit{Opt. Express} \textbf{15,} 15377--15386 (2007).

\bibitem{hentschel2009three} Hentschel, M., Hübel, H., Poppe, A. \& Zeilinger, A. Three-color Sagnac source of polarization-entangled photon pairs. \textit{Opt. Express} \textbf{17,} 23153--23159 (2009).

\bibitem{suhara2009quasi} Suhara, T., Nakaya, G., Kawashima, J. \& Fujimura, M. Quasi-phase-matched waveguide devices for generation of postselection-free polarization-entangled twin photons. \textit{IEEE Photonics Technol. Lett.} \textbf{15,} 1096--1098 (2009).

\bibitem{herrmann2013post} Herrmann, H. \emph{et al}. Post-selection free, integrated optical source of non-degenerate, polarization entangled photon pairs. \textit{Opt. Express} \textbf{21,} 27981--27991 (2013).

\bibitem{laudenbach2016qpmoptics} Laudenbach, F., Hübel, H., Hentschel, M. \& Poppe, A. QPMoptics: a novel tool to simulate and optimise photon pair creation. In: \textit{SPIE Photonics Europe} (pp. 98940V). International Society for Optics and Photonics (2016).

\bibitem{fan1987second} Fan, T. Y. \emph{et al}. Second harmonic generation and accurate index of refraction measurements in flux-grown KTiOPO 4. \textit{Appl. Opt.} \textbf{26,} 2390--2394 (1987).

\bibitem{emanueli2003temperature} Emanueli, S. \& Arie, A. Temperature-dependent dispersion equations for KTiOPO4 and KTiOAsO4. \textit{Appl. Opt.} \textbf{42,} 6661--6665 (2003).

\bibitem{fradkin1999tunable} Fradkin, K., Arie, A., Skliar, A. \& Rosenman, G. Tunable midinfrared source by difference frequency generation in bulk periodically poled KTiOPO4. \textit{Appl. Phys. Lett.} \textbf{74,} 914--916 (1999).

\bibitem{clauser1969proposed} Clauser, J. F., Horne, M. A., Shimony, A. \& Holt, R. A. Proposed experiment to test local hidden-variable theories. \textit{Phys. Rev. Lett.} \textbf{23,} 880 (1969).

\end{thebibliography}
\end{document}